\documentclass[aps,prl,reprint,twocolumn,amsfonts,amssymb,amsmath,superscriptaddress,floatfix,tightenlines]{revtex4-1}

\usepackage{graphicx}
\usepackage{pdfpages}
\usepackage{relsize}
\usepackage[colorlinks=true,citecolor=blue,urlcolor=blue]{hyperref}

\makeatletter
\AtBeginDocument{\let\LS@rot\@undefined}
\makeatother

\def\p{\partial}
\def\mb{\mathbf}

\def\FF{\mathrm{FF}}

\begin{document}

\title{Collective mode contributions to the Meissner effect: 
Fulde-Ferrell and pair-density wave superfluids}

\author{Rufus Boyack}
\affiliation{James Franck Institute, University of Chicago, Chicago, Illinois 60637, USA}
\author{Chien-Te Wu}
\affiliation{James Franck Institute, University of Chicago, Chicago, Illinois 60637, USA}
\affiliation{Department of Electrophysics, National Chiao Tung University, Hsinchu 30010, Taiwan, Republic of China}
\author{Brandon M. Anderson}
\affiliation{James Franck Institute, University of Chicago, Chicago, Illinois 60637, USA}
\author{K. Levin}
\affiliation{James Franck Institute, University of Chicago, Chicago, Illinois 60637, USA}

\begin{abstract}
In this paper we demonstrate the necessity of including the generally omitted collective mode contributions
in calculations of the Meissner effect for non-uniform superconductors. 
We consider superconducting pairing with non-zero center of mass momentum, 
as is relevant to high transition temperature cuprates, cold atoms, and quantum chromodynamic superconductors.
For the concrete example of the Fulde-Ferrell phase we present a quantitative calculation
of the superfluid density, showing the collective mode contributions
are not only appreciable but that they derive from the amplitude mode of the order parameter.
This latter mode (related to the Higgs mode in a charged system)
is generally viewed as being invisible in conventional superconductors.
However, our analysis shows that it is extremely important in pair-density wave type superconductors,
where it destroys superfluidity well before the mean-field order parameter vanishes.
\end{abstract}

\maketitle

There is a strong current interest in collective modes in superconductors in large
part stimulated by the excitement surrounding the discovery of the Higgs boson \cite{PhysicsToday}.
Nevertheless, there is a widespread belief that observing these modes, 
directly or indirectly, is particularly challenging \cite{Parks1969, Varma_review_2015}.
As a result they only infrequently appear in condensed matter 
physics \cite{Littlewood&Varma_1981,Browne&Levin_1983,Endres_2012,Pollet_2012,Auerbach_2014,Brandon_pathintegral_2016,Matsunaga_2013}.
In this paper we show that in a class of very topical superconductors,
collective mode effects associated with the amplitude of the order parameter
play an essential role in the most fundamental quantity, the superfluid density tensor $n_{s}^{ij}$.
The superconductors in question are those which have a ``pair-density wave" order parameter.
These are a large class of superconductors exhibiting
pairing of electrons at non-zero center of mass momentum ${\bf{Q}}$.
Much attention has focused on these systems from the perspective of
high temperature superconductivity (in condensed matter physics \cite{Fradkin_2015,SeamusDavis_2016})
and quantum chromodynamics (in particle physics \cite{Casalbuoni_2004}).

For this class of superfluids, the collective mode contribution to the superfluid density 
has been largely ignored in previous literature \cite{Torma_FF_2014,Samokhin_FFLO_2010},
with the exception of the original calculation of the electromagnetic current
by Larkin and Ovchinnikov \cite{Larkin&Ovchinnikov_FFLO_1965}.
Discussion of this effect can also be found in Ref. \cite{Millis_1987} for a different situation involving non $s$-wave superconductors.
In both cases the size and specific nature of the collective mode contributions was not accessible.  

We provide two different, but related, 
derivations of the superfluid density for the tractable case 
of the Fulde-Ferrell (FF) superfluid \cite{Fulde&Ferrell_1964}.
Importantly, this enables us to compute numerical values for the sizeable collective mode effects in $n_{s}^{ij}$.
The first method is based on using the Ward-Takahashi identity in
the Kubo formalism, while the second method is based on studying the equilibrium current.
In both approaches particle number is manifestly conserved and gauge invariance is maintained. 
Through the first approach we find that amplitude collective modes drive the superfluid 
density (along the direction parallel to $\mathbf{Q}$)
to zero at temperatures lower than those associated with the vanishing of the mean-field order parameter.

Before giving these more complete calculations,
here we provide a general argument for the necessity of including collective mode effects in non-uniform superconductors.  
The origin of collective mode contributions to the Meissner effect \cite{Larkin&Ovchinnikov_FFLO_1965,Millis_1987}
lies in the fact \cite{AGD_1975} that, in the presence of a vector potential $A^{\mu}$,
the order parameter $\Delta$ will depend on $A^{\mu}$ through the gap (saddle-point) equation \cite{Brandon_pathintegral_2016}. 
A series expansion of $\Delta[A]$, in powers of $A^{\mu}$, is thus
\begin{equation}
\Delta[A]=\Delta^{(0)}[A=0]+\Delta^{(1)}[A]+\mathcal{O}(A^{2}).
\end{equation}
Here $\Delta^{(0)}$ is the order parameter in the absence of $A^{\mu}$ and $\Delta^{(1)}$ is a correction linear in $A^{\mu}$.
It is this term which gives rise to the rarely discussed collective mode contributions to the superfluid density.
Since $\Delta^{(1)}$ is a scalar quantity, it can depend on only scalar, linear functions of $A^{\mu}$.
Therefore, in a uniform superfluid $\Delta^{(1)}$ is a function of only $\nabla\cdot{\bf A}$ \footnote{In general,
for a uniform superfluid $\Delta^{(1)}$ will be a function of $\p_{\mu}A^{\mu}$. 
For the superfluid density, however, only the vector component of $A^{\mu}$ is of interest.}.
Thus, if one chooses the (``transverse'') gauge such that $\nabla\cdot{\bf A}=0$,
the collective mode contribution $\Delta^{(1)}$ vanishes identically \cite{AGD_1975}. 

However, for a non-uniform system there are other scalar, linear functions of $A^{\mu}$.
In particular, for a pair-density wave superfluid with pairing vector ${\bf Q}$, 
$\Delta^{(1)}$ can depend on other scalar, linear quantities such as $\mb{A}\cdot\mb{Q}$.
Hence, for this non-uniform superfluid, even in the gauge where $\nabla\cdot{\bf A}=0$,
$\Delta^{(1)}$ may still be non-zero.
In principle, this allows for a collective mode contribution to the superfluid density.
[For future use in the discussion below,
we define $\Delta^{(1)}=\left.(d\Delta[A]/dA_{\mu})\right|_{A=0}A_{\mu}=\int dq\ \Pi^{\mu}(q)A_{\mu}(q)$.]

To illustrate this effect, we implement this argument for the specific case of the Fulde-Ferrell \cite{Fulde&Ferrell_1964} superfluid.
For simplicity the FF pairing vector is assumed to be $\mathbf{Q}=Q\hat{z}$.
In the FF phase specifically, both a continuous rotational and global gauge symmetry are spontaneously broken.
Similarly discrete time-reversal symmetry is also spontaneously broken. 
However, gauge invariant observables are translationally invariant \cite{Radzihovsky_FFLO_2011}. 
Due to the underlying rotational symmetry of the FF state,
the superfluid density vanishes along the directions transverse to $\mb{Q}$.
Hence $n_{s}^{xx}=n_{s}^{yy}=0$, and thus only $n_{s}^{zz}$ needs to be considered \cite{Radzihovsky_FFLO_2011}.
As has been posited \cite{Fulde&Ferrell_1964}, and will be shown in more detail below, the superfluid density can be written as 
\begin{equation}\label{eq:SFD0}
\left.\frac{\p j^{z}}{\p Q}\right|_{\mu,h}=\frac{1}{2}\left(\frac{n^{zz}_{s}}{m}\right),
\end{equation}
where $j^{z}(Q)$ is the equilibrium current. 

It is useful to express Eq.~(\ref{eq:SFD0}) in terms of the mean-field thermodynamic potential $\Omega$,
where $j^{z}(Q)=2\left.(\p\Omega/\p Q)\right|_{\mu,h,\Delta}$.
The mean-field values of the chemical potential, gap, magnetic field, and pairing vector 
are denoted by $\mu_{0}, \Delta_{0}, h_{0},$ and $Q_{0}$ respectively.
The saddle-point condition which determines $Q_{0}$ is then $j^{z}(Q_{0})=0$. 
Similarly, the saddle-point condition which determines $\Delta_{0}$ is
$\left.(\p\Omega/\p\Delta)\right|_{\mu,h,Q}=0$. 
In terms of $\Omega$, Eq.~(\ref{eq:SFD0}) becomes
\begin{equation}\label{eq:djdQ0}
\left.\frac{\p j^{z}}{\p Q}\right|_{\mu,h}=
2\left[\left.\frac{\p^{2} \Omega}{\partial Q^2}\right|_{\mu,h,\Delta}
-\left(\frac{\p^{2}\Omega}{\p\Delta\p Q}\right)^{2}
\mathlarger{\mathlarger{/}}\left.\frac{\p^{2}\Omega}{\p\Delta^{2}}\right|_{\mu,h,Q}\right],
\end{equation}
where both saddle-point equations, and the symmetry of mixed partial derivatives has been used. 

Equation (\ref{eq:djdQ0}) indicates that there are two contributions to the superfluid density.
The first is the conventional ``bubble" term (which is usually assumed to
be sufficient) and the second represents the collective mode contribution required for gauge invariance. 
Importantly, a stability inequality for the FF superfluid based on the thermodynamic potential curvature \cite{ChihChun_FFLO_2006,Yan_FFLO_2007,Qijin_FFLO_2017}
is equivalent to requiring that both $n^{zz}_{s}$, as derived above, and
$\left.\left(\p^{2}\Omega/\p\Delta^{2}\right)\right|_{\mu,h,Q}$ are positive.
From this, it follows that for a stable FF superfluid 
the collective mode contribution always acts to reduce the overall size of the superfluid density.
The above arguments, however, still do not indicate how large the magnitude of this effect is.

In this paper the collective-mode contribution will be shown to be appreciable;
this underlines the inadequacy of including only the so-called
bubble term \cite{Torma_FF_2014,Samokhin_FFLO_2010}.
Equally important is the nature of these collective mode corrections.
For the FF superfluid we will show that they derive
from the \textit{amplitude} mode of the order parameter.
This mode is thought to be rather invisible in conventional superconductors \cite{Parks1969}.
Nevertheless we demonstrate how it arises to ensure the electromagnetic (EM) response is manifestly gauge invariant.
In this way, it importantly affects the superfluid density in pair-density wave superconductors.
Readers uninterested in the technical details can skip to the numerical results for a 
simple understanding of our main results. 

\textit{Mean-field formalism.--}
The FF mean-field Hamiltonian, in the $\psi^{\mathsf{T}}_{\mb{k}}~=(c_{\mb{k},\uparrow},c_{\mb{-k+Q},\downarrow}^{\dagger})$ basis, is 
$H_{\FF}~=\sum_{\mb{k}}\psi_{\mb{k}}^{\dagger}\mathcal{H}_{\FF}\psi_{\mb{k}}$, 
where \cite{Qijin_FFLO_2007}
\begin{equation}
\mathcal{H}_{\FF}=\left(\begin{array}{cc}\xi_{\mb{k},\uparrow}&-\Delta\\-\Delta^{*}&-\xi_{\mb{k-Q},\downarrow}.
\end{array}\right).
\end{equation}
Here an irrelevant constant $-\sum_{\mb{k}}\xi_{\mb{k-Q},\downarrow}$ has been ignored.
The notation is as follows: the dispersion relation is defined by $\xi_{\mb{k},\sigma}=\mb{k}^2/2m-\mu_{\sigma}$,
where $\mu_{\sigma}$ is the fermionic chemical potential for a species with spin $\sigma=\uparrow,\downarrow$,
$m$ is the fermion mass, and $\Delta$ denotes an $s$-wave pairing gap. 
It is useful to define
$\mu=\tfrac{1}{2}\left(\mu_{\uparrow}+\mu_{\downarrow}\right)$ and
$h=\tfrac{1}{2}\left(\mu_{\uparrow}-\mu_{\downarrow}\right)$.
The dispersion relations are then written compactly as
$\xi_{\mb{kQ}}=(1/2m)[\mb{k}^2+(\mb{Q}/2)^2]-\mu$, $E_{\mb{kQ}}^2=\xi_{\mb{kQ}}^2+\Delta^2$,
$h_{\mb{kQ}}=h-\mb{k}\cdot{\mb{Q}}/2m$. Throughout the paper $\hbar=k_{B}=c=1$.

The inverse Nambu Green's function is then $\mathcal{G}^{-1}=i\omega_{n}-\mathcal{H}_{\FF},$
where $i\omega_{n}$ is a fermionic Matsubara frequency. 
The inverse bare Green's function is defined by $G^{-1}_{0,\sigma}(k)=i\omega_{n}-\xi_{\mb{k}}$.
Thus, the off diagonal Gorkov function is
$\mathcal{G}_{12}(k)=\Delta G_{0,\downarrow}(-k+Q)G_{\uparrow}(k)$,
where the (spin-up) Green's function is $G_{\uparrow}(k)=\mathcal{G}_{11}(k)$. 
Note that Greek indices denote spacetime coordinates: $\mu=(t,x,y,z)$; 
whereas Roman indices denote spatial coordinates: $i=(x,y,z)$. 
Here $Q^{\mu}=(0,\mb{Q})$.
Explicit calculation then gives the full Green's function
which has appeared in the literature \cite{Qijin_FFLO_2007,Torma_FF_2014}.
Using this, the particle number is $n=\sum_{\sigma}\sum_{k}G_{\sigma}(k)$,
where $\sum_{k}\equiv \beta^{-1}\sum_{i\omega_{n}}\sum_{\mb{k}}$ 
with $\beta$ being inverse temperature.

From Dyson's equation, $G^{-1}_{\sigma}(k)=G^{-1}_{0,\sigma}(k)-\Sigma_{\sigma}(k),$
the self energy is $\Sigma_{\sigma}(k)=-|\Delta|^2G_{0,\bar{\sigma}}(-k+Q)$.
An important identity is then
$\mathcal{G}_{12}(\widetilde{k})=\Delta G_{0,\downarrow}(-\widetilde{k}_{-})G_{\uparrow}(\widetilde{k}_{+})
=\Delta G_{0,\uparrow}(\widetilde{k}_{+})G_{\downarrow}(-\widetilde{k}_{-})$.
[For convenience, we have defined $\widetilde{k}^{\mu}_{\pm}\equiv k^{\mu}\pm Q^{\mu}/2$.]

We now study the EM response of this superfluid.
For the issue of primary concern in this paper (the superfluid density)
the distinction between neutral and charged superfluids is irrelevant. 
For the purposes of simplicity, our general equations are for neutral superfluids.
We apply linear response theory, where, a fictitious vector potential $A^{\mu}$ is applied,
and at the end of the calculation $A^{\mu}\rightarrow0$.
The EM current is $j^{\mu}(q)=K^{\mu\nu}(q)A_{\nu}(q)$, where $K^{\mu\nu}(q)$ is the EM response kernel.
The response kernel can also be expressed as $K^{\mu\nu}(q)=P^{\mu\nu}(q)+(n/m)\delta^{\mu\nu}(1-\delta_{\mu,0})$
(with $\mu$ and $\nu$ not summed over) where the EM response functions are denoted by $P^{\mu\nu}(q)$.
In the Kubo formalism the EM response functions for a superfluid are
\begin{equation}\label{eq:ERF}
P^{\mu\nu}(q)=\sum_{\sigma}\sum_{k}G_{\sigma}(k_{+})\Gamma^{\mu}_{\sigma}(k_{+},k_{-})G_{\sigma}(k_{-})\gamma^{\nu}_{\sigma}(k_{-},k_{+}).
\end{equation}
Here $q^{\mu}=(i\Omega_{m},\mb{q})$, with $i\Omega_{m}$ a bosonic Matsubara frequency.
The quantity $\Gamma^{\mu}(k_{+},k_{-})$ denotes the full EM vertex,
where the incoming (outgoing) momentum is $k_{+}$ $(k_{-}$), with $k^{\mu}_{\pm}\equiv k^{\mu}\pm q^{\mu}/2$ .
To determine the full vertex $\Gamma^{\mu}(k_{+},k_{-})$, we apply the Ward-Takahashi identity (WTI) \cite{Ryder}:
$q_{\mu}\Gamma^{\mu}_{\sigma}(k_{+},k_{-})=G^{-1}_{\sigma}(k_{+})-G^{-1}_{\sigma}(k_{-})
=q_{\mu}\gamma^{\mu}_{\sigma}(k_{+},k_{-})+\Sigma_{\sigma}(k_{-})-\Sigma_{\sigma}(k_{+}).$
This is an exact relation in quantum field theory
which relates the single particle Green's function to the full vertex. 
It is a gauge invariant statement and here it reflects the 
underlying global $U(1)$ gauge symmetry. 
Furthermore, satisfying the WTI ensures conservation of particle number
(or charge in the charged superfluid case).
The bare WTI, $q_{\mu}\gamma^{\mu}_{\sigma}(k_{+},k_{-})=G^{-1}_{0,\sigma}(k_{+})-G^{-1}_{0,\sigma}(k_{-})$,
is satisfied by the bare vertex $\gamma^{\mu}_{\sigma}(k_{+},k_{-})=(1,\textbf{k}/m)$.

In the limit $q^{\mu}\rightarrow0$, the WTI reduces to the Ward identity:
$\Gamma^{\mu}_{\sigma}(k,k)=\gamma^{\mu}_{\sigma}(k,k)-(\p\Sigma_{\sigma}(k)/\p k_{\mu})$.
The second term diagrammatically represents a vertex insertion in the self energy. 
This relation then importantly shows that the full vertex can be obtained 
by performing all possible vertex insertions in the full Green's function \cite{Ryder, Nambu}.
Due to the spontaneous symmetry breaking of the global $U(1)$ symmetry,
a collective mode vertex $\Pi^{\mu}(q)$ $(\bar{\Pi}^{\mu}(q))$ can be inserted
into the order parameter $\Delta$ $(\Delta^{*})$. 
After these vertex insertions, the full vertex can be determined exactly.
Importantly, the collective mode effects enter in the superfluid density via the combination
$\left(\Delta^{*}\Pi^{\mu}(0)+\Delta\bar{\Pi}^{\mu}(0)\right)$. 

Now we show that it is the amplitude rather than phase mode which
is important to the calculation of the FF superfluid density.
The collective mode vertices are self-consistently determined from the gap equation and, 
for $q^{\mu}\neq0$, satisfy $q_{\mu}\Pi^{\mu}(q)=2\Delta, q_{\mu}\bar{\Pi}^{\mu}(q)=-2\Delta^{*}$. 
Note that, as $q^{\mu}\rightarrow0$, the right hand side of these expressions remains finite,  
and so must the left hand side. 
It follows that $\Pi^{\mu}(q)$ and $\bar{\Pi}^{\mu}(q)$ are singular in the $q^{\mu}\rightarrow0$ limit.
This pole corresponds to the phase mode of the gap; 
equivalently it reflects the Nambu-Goldstone boson associated with 
the spontaneous symmetry breaking of global $U(1)$.
On the other hand, note that 
$q_{\mu}\left(\Delta^{*}\Pi^{\mu}(q)+\Delta\bar{\Pi}^{\mu}(q)\right)=0$. 
This identity is non-singular in the $q^{\mu}\rightarrow0$ limit,
and reflects the fact that $\Delta^{*}\Pi^{\mu}(q)+\Delta\bar{\Pi}^{\mu}(q)$
does not have a zero momentum pole.
Thus, this quantity corresponds to the amplitude mode of the gap. 

The gap equation for an FF superfluid is 
$\Delta/g=\Delta\sum_{\sigma}\sum_{k}G_{0,\downarrow}(-k+Q)G_{\uparrow}(k)=\sum_{\sigma}\sum_{k}\mathcal{G}_{12}(\widetilde{k})$ \cite{Qijin_FFLO_2007}.
Expressing this equation diagrammatically
allows one to perform all vertex insertions on both the gaps ($\Delta$, $\Delta^{*}$)
and on the bare and full Green's functions. This procedure then leads to 
the explicit form of the collective mode vertices $\Pi^{\mu}, \bar{\Pi}^{\mu}$,
as shown explicitly in the Supplemental Material.
From this analysis one can then obtain the zero momentum limit 
of the quantity $\Delta^{*}\Pi^{z}+\Delta\bar{\Pi}^{z}$ of importance here.

Note that $\Delta$ is a function of the FF pairing vector $Q$,
and by differentiating the gap equation with respect to $Q$ (at fixed $\mu$ and $h$)
one can obtain $\left.(\p|\Delta|^2/\p Q)\right|_{\mu,h}$.
An explicit calculation then gives the following important identity (for $\Delta\neq0$):
\begin{equation}\label{eq:CME1}
\Delta^{*}\Pi^{z}(0)+\Delta\bar{\Pi}^{z}(0)=P^{z}_{0}/M_{0}=2\left.(\p|\Delta|^2/\p Q)\right|_{\mu,h}.
\end{equation}
The order of limits in which frequency and momentum are taken to zero is important;
frequency $i\Omega_{m}$ and $q^{z}$ are set to zero, and then $q^{x},q^{y}\rightarrow0$.
In the following section this will be clarified.
The quantities $P^{z}_{0}$ and $M_{0}$ are generalized three-particle and four-particle 
Green's functions, respectively, which are defined in the next section.
The generalized Green's functions in Eq. (\ref{eq:CME1}) also appear in a similar form in the work of 
Larkin and Ovchinnikov \cite{Larkin&Ovchinnikov_FFLO_1965} and Millis \cite{Millis_1987}.
Finally, note that when $Q=0$, $P^{z}_{0}=0$, and thus this collective mode term does not contribute 
for a homogeneous superfluid. 

\textit{Superfluid density derivation via Kubo formula.--}
In this section we use the Kubo formula and Eq.~(\ref{eq:ERF}) 
to derive the superfluid density tensor:
\begin{equation}\label{eq:SFD1}
(n^{ij}_{s}/m)=(n/m)\delta^{ij}+P^{ij}(\omega=0,\mb{q}\rightarrow0).
\end{equation}
Note that, the order of limits in the above expression is crucial.
To compute $n^{ij}_{s}$, first set $\omega=q^{i}=q^{j}=0$, then take $q^{k}\rightarrow0$, where  $k\neq i,j$.
The collective modes are contained within the second term.

Evaluating this expression we find 
\begin{align}\label{eq:SFD2}
\left(\frac{n^{ij}_{s}}{m}\right)&=\sum_{\mb{k}}\frac{\Delta^2}{E^2_{\mb{kQ}}}\left(\frac{X_{\mb{k}}}{E_{\mb{kQ}}}-\beta Y_{\mb{k}}\right)(\widetilde{k}^{i}_{-}/m)(\widetilde{k}^{j}_{+}/m)\nonumber\\&-\delta^{iz}\delta^{jz}(P^{z}_{0})^2/M_{0}.
\end{align}
where we define $X_{\mb{k}}\equiv D^{-1}\sinh(\beta E_{\mb{k Q}})$, 
and $Y_{\mb{k}}\equiv D^{-2}(1+\cosh(\beta E_{\mb{k Q}})\cosh(\beta h_{\mb{k Q}}))$
with $D\equiv\cosh(\beta E_{\mb{k Q}})+\cosh(\beta h_{\mb{k Q}})$.

The first term in Eq.~(\ref{eq:SFD2}) represents the usual \cite{Torma_FF_2014,Samokhin_FFLO_2010}
``bubble" contribution, due to bubble terms in both $(n/m)\delta^{ij}$ and $P^{ij}(0)$.
The second term represents the collective mode contribution arising solely from $P^{ij}(0)$.
As an important check we note that Eq.~(\ref{eq:SFD2}) is identical to Eq.~(\ref{eq:SFD0}) and Eq.~(\ref{eq:djdQ0}), 
where explicit calculation shows that the bubble term is
$4\left.(\p^{2}\Omega/\p Q^2)\right|_{\mu,h,\Delta}$ and $(\p^{2}\Omega/\p\Delta\p Q)=-\Delta P^{z}_{0},
\left.(\p^{2}\Omega/\p\Delta^2)\right|_{\mu,h,Q}=4\Delta^2 M_{0}$.
Here $\Omega=\Delta^2/g-\beta^{-1}\sum_{\mb{k}}\left\{\mathrm{log}[2\cosh(\beta E_{\mb{kQ}})+2\cosh(\beta h_{\mb{kQ}})]-\beta \xi_{\mb{kQ}}\right\}$ 
is the mean-field thermodynamic potential \cite{Torma_FF_2014, Qijin_FFLO_2017}. 

Note that the collective mode contribution is only along the direction of the FF pairing vector,
in agreement with the general arguments presented earlier.
Direct calculation shows that $n^{ij}_{s}$ is diagonal,
with $n^{xx}_{s}=n^{yy}_{s}=0$, as required by symmetry.

\textit{Superfluid density derivation via equilibrium current.--}
A verification of this Kubo analysis and the collective
mode contributions can be made in a slightly simpler fashion.
Here we derive the superfluid density in the direction along the FF pairing vector using
only the equilibrium current and its partial derivative with respect to $Q$. 
The equilibrium current in the $z$-direction is 
$j^{z}(Q)=\sum_{\sigma}\sum_{k}(\widetilde{k}^{z}_{+}/m)G_{\sigma}(\widetilde{k})$.
This expression follows from $j^{z}=2\left.(\p\Omega/\p Q)\right|_{\mu,h,\Delta}$. 
By symmetry the mean-field currents in the other directions vanish: $j^{x}=j^{y}=0$. 

In what follows it will be important to fix $\mu$ and $h$, and to consider
the $Q$-dependence of only the gap:~$\Delta(Q)$. 
The following lemma, whose proof is given in the Supplemental Material, will also be required:
$\left.(\p G^{-1}_{\sigma}(\widetilde{k}_{+})/\p Q)\right|_{\mu,h}
=-(1/2)\Gamma^{z}_{\sigma}(\widetilde{k}_{+},\widetilde{k}_{+}).$
The partial derivative of $j^{z}$ can now be computed.
Using the number equation $n=\sum_{\sigma}\sum_{k}G_{\sigma}(k)$, along with the aforementioned Lemma,
the partial derivative of $j^{z}$ is then
$\left.(\p j^{z}/\p Q)\right|_{\mu,h}
=(n/2m)-\sum_{\sigma}\sum_{k}(\widetilde{k}^{z}_{+}/m)G^{2}_{\sigma}(\widetilde{k})\left.(\p G^{-1}_{\sigma}(\widetilde{k}_{+})/\p Q)\right|_{\mu,h}
=(n^{zz}_{s}/2m)$.
Note that the above expression, which reproduces Eq.~(\ref{eq:SFD0}) and Eq.~(\ref{eq:SFD2}),
includes collective mode contributions arising through $\Gamma^{z}(\widetilde{k}_{+},\widetilde{k}_{+})$. 

\begin{figure}[t]
\includegraphics[width=8cm]{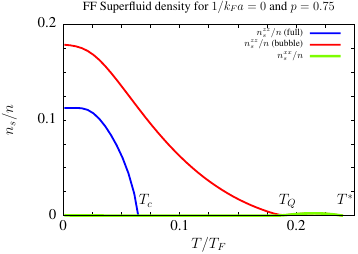}
\caption{Superfluid density as a function of temperature for the FF phase at unitarity. 
The polarization $p=(n_{\uparrow}-n_{\downarrow})/n$ is set to $p=0.75$
and the inverse scattering length is $1/k_{F}a=0$.
The blue curve is the full expression for $n^{zz}_{s}/n$ while the red curve
is the bubble contribution alone. The green curve is $n^{xx}_{s}/n$;
in this case there are no collective modes.}
\label{fig:FF_rhos}
\end{figure}

\textit{Numerical Results.--}
In Fig.~(\ref{fig:FF_rhos}) the superfluid density with collective mode effects (blue curve) is plotted as a function of temperature 
for the case of a polarized superfluid with $p=0.75$ and interaction strength (via the scattering amplitude) $1/k_{F}a=0$. 
These parameters were chosen because there are independent calculations in the literature \cite{Qijin_FFLO_2017} 
claiming to establish a regime of stability for this FF phase.
There it was argued that the thermodynamic potential must satisfy two conditions: $\left.(\p^{2}\Omega/\p\Delta^{2})\right|_{\mu,h,Q}>0$
as well as the condition requiring that the right hand side of Eq.~(\ref{eq:djdQ0}) is positive.
Importantly, this latter stability criterion is precisely equivalent to $n^{zz}_{s}$ being positive.
Our independent calculations yield a largest temperature for which the superfluid is stable to be $T_c/T_{F}\sim0.6-0.65$, 
in rough agreement with Ref. \cite{Qijin_FFLO_2017} for the same input parameters.
It should be noted that, although we are in the strong interaction regime, 
for quantitative purposes strict mean-field parameters are used in these plots.  
For numerical checks we have verified that our mean-field solutions are
global minima of the thermodynamic potential \cite{Radzihovsky&Sheehy_2010}
and that the blue curve computed via Eq. (\ref{eq:SFD2}) is numerically
equivalent to that computed via the equilibrium current using Eq.~(\ref{eq:SFD0}).

The red curve in Fig.~(\ref{fig:FF_rhos}) is the bubble contribution which is usually 
\cite{Torma_FF_2014,Samokhin_FFLO_2010} all that is considered.
The green curve plots the transverse superfluid density.
As required by symmetry, $n^{xx}_{s}=0$ for all $T<T_{Q}$ for which the FF pairing vector $Q$ persists.
When $T\geq T_Q\sim0.2T_{F}$, the FF phase ceases to exist and instead a Sarma superfluid (with $Q\equiv 0$) exists. 
In this regime $n^{zz}_{s}$ has no collective mode contribution so
that $n^{zz}_{s}=n^{xx}_{s}=n^{yy}_{s}$.
This explains the small regime above $T_Q$
where $n^{xx}_{s}>0$, which persists until $\Delta=0$ at $T^*$. 
It can be noted that the effects of the collective modes are quite appreciable in this plot.
This follows because the bubble term is proportional to $(\Delta/E_{F})^2$,
whereas the collective mode term is proportional to $(Q/k_{F})^2$.
(Note though the integrands in both expressions are somewhat different.) 
Near zero temperature, with $p=0.75$ and $1/k_{F}a=0$,
$\Delta/E_{F}\sim0.16$ whereas $Q/k_{F}\sim0.71$.
Thus qualitatively the collective mode contribution is expected to be an important contribution.

This figure encapsulates the important point that collective modes of the
order parameter can substantially reduce the transition temperature from its mean-field value.
In this way a pairing gap persists above the temperature at which the superfluid density disappears. 
This is a variant on a particular pseudogap scenario in the high $T_c$ cuprates \cite{Emery_1995}
which invokes phase rather than amplitude collective mode contributions to suppress $T_{c}$ relative to its mean field value.

\textit{Conclusions.--}
In this paper we have computed the superfluid density tensor $n^{ij}_{s}$ for the FF superfluid phase. 
Importantly, we have shown (using multiple, distinct theoretical frameworks) 
that widely neglected collective (amplitude) mode contributions cannot be ignored.
In general they will affect $n^{ij}_{s}$ for the broad class of $\mathbf{Q}\neq 0$ pair-density wave superconductors.
Indeed, while Fig.~(\ref{fig:FF_rhos}) was obtained using the specific microscopic approach of Fulde and Ferrell, 
we believe its qualitative features (except for the behavior of the transverse superfluid density) are more generic.
This figure suggests that a conventional Landau Ginsburg expansion \cite{Larkin&Ovchinnikov_FFLO_1965} may be problematic.
In a temperature range near but slightly above $T_c$, the mean-field gap persists so that $\Delta$ may not be sufficiently small.
Similarly, at temperatures near $T^*$ where this parameter is appropriately small,
the stable phase has vanishing $\mathbf{Q}$.
This poses a challenge for future work.

Given the intense interest in condensed matter observations of a Higgs mode,
one can inquire as to what is the relation between the amplitude mode
evident in pair-density wave superconductors and the Higgs mode in condensed matter
\cite{Littlewood&Varma_1981,Browne&Levin_1983,Auerbach_2014,Brandon_pathintegral_2016,Endres_2012,Matsunaga_2013}.
The Higgs mechanism is associated with a charged system and the present theory of the superfluid density, 
which depends on the amplitude mode, is applicable to both charged and neutral superfluids. 
However, because we are considering the case of zero frequency and zero wave number 
the Higgs mode is not observed as a collective resonance.
Nevertheless, we have ascertained in this paper that its very existence has important
consequences for readily accessible physical quantities in pair-density wave superconductors.

\textit{Acknowledgements.--}
We thank Qijin Chen and Matthew Roberts for helpful conversations.
This work was supported by NSF-DMR-MRSEC 1420709.

\bibliography{Review}

\clearpage


\section*{Supplementary material}

# Supplemental Material: Collective mode contributions to the Meissner effect: Fulde-Ferrell and pair-density wave superfluids

Rufus Boyack[1], Chien-Te Wu[1,2], Brandon M. Anderson[1], and K. Levin[1]

[1]James Franck Institute, University of Chicago, Chicago, Illinois 60637, USA and

[2]Department of Electrophysics, National Chiao Tung University, Hsinchu 30010, Taiwan, Republic of China

## I. MEAN-FIELD FORMALISM

### A. Green's functions

The starting point for our study of Fulde-Ferrell superfluids is the mean-field Hamiltonian [1]:

$$H_{\mathrm{FF}} = \sum_{\mathbf{k}} \left[ \xi_{\mathbf{k},\uparrow} c^{\dagger}_{\mathbf{k},\uparrow} c_{\mathbf{k},\uparrow} + \xi_{\mathbf{k}-\mathbf{Q},\downarrow} c^{\dagger}_{-\mathbf{k}+\mathbf{Q},\downarrow} c_{-\mathbf{k}+\mathbf{Q},\downarrow} + \Delta c^{\dagger}_{-\mathbf{k}+\mathbf{Q},\downarrow} c^{\dagger}_{\mathbf{k},\uparrow} + \Delta^{*} c_{\mathbf{k},\uparrow} c_{-\mathbf{k}+\mathbf{Q},\downarrow} \right]. \tag{1}$$

The matrix representation of this Hamiltonian, in the $\psi^{\mathsf{T}}_{\mathbf{k}} = (c_{\mathbf{k},\uparrow}, c^{\dagger}_{-\mathbf{k}+\mathbf{Q},\downarrow})$ basis, is $H_{\mathrm{FF}} = \sum_{\mathbf{k}} \psi^{\dagger}_{\mathbf{k}} \mathcal{H}_{\mathrm{FF}} \psi_{\mathbf{k}}$, where

$$\mathcal{H}_{\mathrm{FF}} = \begin{pmatrix} \xi_{\mathbf{k},\uparrow} & -\Delta \\ -\Delta^{*} & -\xi_{\mathbf{k}-\mathbf{Q},\downarrow} \end{pmatrix}. \tag{2}$$

Here we have ignored an irrelevant constant $-\sum_{\mathbf{k}} \xi_{\mathbf{k}-\mathbf{Q},\downarrow}$. The inverse Nambu Green's function is then $\mathcal{G}^{-1} = i\omega_n - \mathcal{H}_{\mathrm{FF}}$, where $i\omega_n$ is a fermionic Matsubara frequency. Performing the matrix inverse we obtain

$$\mathcal{G}(k) = \frac{1}{(i\omega_n - \xi_{\mathbf{k},\uparrow})(i\omega_n + \xi_{\mathbf{k}-\mathbf{Q},\downarrow}) - |\Delta|^2} \begin{pmatrix} i\omega_n + \xi_{\mathbf{k}-\mathbf{Q},\downarrow} & -\Delta \\ -\Delta^{*} & i\omega_n - \xi_{\mathbf{k},\uparrow} \end{pmatrix}. \tag{3}$$

If we shift $\mathbf{k} \to \mathbf{k} + \mathbf{Q}/2$, then redefine $\mathbf{Q} \to 2\mathbf{Q}$, this Nambu Green's function agrees with Eq. (3) of Ref. [2]. The single particle Green's function can then be found from the above equation. Namely, $G_{\uparrow}(k) = \mathcal{G}_{11}(k)$ and $-G_{\downarrow}(-k+Q) = \mathcal{G}_{22}(k)$. If we perform this inversion, and define the bare Green's function by $G^{-1}_{0,\sigma}(k) = i\omega_n - \xi_{\mathbf{k},\sigma}$, then the full Green's function is

$$G^{-1}_{\sigma}(k) = G^{-1}_{0,\sigma}(k) - \Sigma_{\sigma}(k), \tag{4}$$

where the self energy is given by

$$\Sigma_{\sigma}(k) = -|\Delta|^2 G_{0,\bar{\sigma}}(-k+Q) = \frac{|\Delta|^2}{i\omega_n + \xi_{\mathbf{k}-\mathbf{Q},\bar{\sigma}}}. \tag{5}$$

Here $k^{\mu} = (i\omega_n, \mathbf{k})$, $Q^{\mu} = (0, \mathbf{Q})$ and without loss of generality the FF pairing vector is assumed to be along the $\hat{z}$-direction: $\mathbf{Q} = Q\hat{z}$. The full Green's function is thus

$$G_{\sigma}(k) = \frac{(i\omega_n + \xi_{\mathbf{k}-\mathbf{Q},\sigma})}{(i\omega_n - \xi_{\mathbf{k},\sigma})(i\omega_n + \xi_{\mathbf{k}-\mathbf{Q},\sigma}) - |\Delta|^2}. \tag{6}$$

Similarly, from the Nambu Green's function in Eq. (3) one has $\mathcal{G}_{12}(k) = \Delta G_{0,\downarrow}(-k+Q) G_{\uparrow}(k) = \Delta G_{0,\uparrow}(k) G_{\downarrow}(-k+Q)$. It will be convenient later to express the Gorkov function symmetrically by

$$\mathcal{G}_{12}(k+Q/2) = \Delta G_{0,\downarrow}(-k+Q/2) G_{\uparrow}(k+Q/2) = \Delta G_{0,\uparrow}(k+Q/2) G_{\downarrow}(-k+Q/2). \tag{7}$$

For a polarized superfluid it is useful to define the chemical potential and effective magnetic field by

$$\mu = \frac{1}{2}(\mu_{\uparrow} + \mu_{\downarrow}), \tag{8}$$

$$h = \frac{1}{2}(\mu_{\uparrow} - \mu_{\downarrow}). \tag{9}$$

For convenience we shall need the following dispersion relations

$$\xi_{\mathbf{kQ}} = \frac{1}{2m}\left(\mathbf{k}^2 + (\mathbf{Q}/2)^2\right) - \mu,$$ (10)

$$E_{\mathbf{kQ}}^2 = \xi_{\mathbf{kQ}}^2 + \Delta^2,$$ (11)

$$h_{\mathbf{kQ}} = h - \frac{\mathbf{k}\cdot\mathbf{Q}/2}{m},$$ (12)

$$g_{\mathbf{kQ}} = h + \frac{\mathbf{k}\cdot\mathbf{Q}/2}{m} = h_{-\mathbf{k},\mathbf{Q}}.$$ (13)

Our primary interest is for the case where $\Delta = \Delta^*$, and here it has been assumed $\Delta$ is real. The coherence factors of the Green's functions are then defined by

$$u_{\mathbf{kQ}}^2 = \frac{1}{2}\left(1 + \frac{\xi_{\mathbf{kQ}}}{E_{\mathbf{kQ}}}\right),$$ (14)

$$v_{\mathbf{kQ}}^2 = \frac{1}{2}\left(1 - \frac{\xi_{\mathbf{kQ}}}{E_{\mathbf{kQ}}}\right).$$ (15)

Similarly the poles of the Green's functions are defined by

$$x_{1,\uparrow} = E_{\mathbf{kQ}} - h_{\mathbf{kQ}},$$ (16)

$$x_{2,\uparrow} = E_{\mathbf{kQ}} + h_{\mathbf{kQ}},$$ (17)

$$x_{1,\downarrow} = E_{\mathbf{kQ}} + g_{\mathbf{kQ}},$$ (18)

$$x_{2,\downarrow} = E_{\mathbf{kQ}} - g_{\mathbf{kQ}}.$$ (19)

With the above definitions, it follows that the spin up and spin down Green's functions are

$$G_\uparrow(k + Q/2) = \frac{u_{\mathbf{kQ}}^2}{i\omega_n - x_{1,\uparrow}} + \frac{v_{\mathbf{kQ}}^2}{i\omega_n + x_{2,\uparrow}},$$ (20)

$$G_\downarrow(k + Q/2) = \frac{u_{\mathbf{kQ}}^2}{i\omega_n - x_{1,\downarrow}} + \frac{v_{\mathbf{kQ}}^2}{i\omega_n + x_{2,\downarrow}}.$$ (21)

In terms of the coherence factors, the Gorkov function is

$$\mathcal{G}_{12}(k + Q/2) = -u_{\mathbf{kQ}}v_{\mathbf{kQ}}\left(\frac{1}{i\omega_n - x_{1,\uparrow}} - \frac{1}{i\omega_n + x_{2,\uparrow}}\right).$$ (22)

Finally, it will prove convenient when evaluating the superfluid density to define the following the expressions

$$W_{\mathbf{k}} \equiv \frac{\sinh(\beta h_{\mathbf{kQ}})}{\cosh(\beta E_{\mathbf{kQ}}) + \cosh(\beta h_{\mathbf{kQ}})},$$ (23)

$$X_{\mathbf{k}} \equiv \frac{\sinh(\beta E_{\mathbf{kQ}})}{\cosh(\beta E_{\mathbf{kQ}}) + \cosh(\beta h_{\mathbf{kQ}})},$$ (24)

$$Y_{\mathbf{k}} \equiv \frac{1 + \cosh(\beta E_{\mathbf{kQ}})\cosh(\beta h_{\mathbf{kQ}})}{[\cosh(\beta E_{\mathbf{kQ}}) + \cosh(\beta h_{\mathbf{kQ}})]^2},$$ (25)

$$Z_{\mathbf{k}} \equiv \frac{\sinh(\beta E_{\mathbf{kQ}})\sinh(\beta h_{\mathbf{kQ}})}{[\cosh(\beta E_{\mathbf{kQ}}) + \cosh(\beta h_{\mathbf{kQ}})]^2}.$$ (26)

The $X_{\mathbf{k}}$ and $Y_{\mathbf{k}}$ expressions are the same definitions as given in Ref. [2], however, the $Z_{\mathbf{k}}$ expression is slightly modified. The $W_{\mathbf{k}}$ expression is our own definition.

## B. Mean-field equations

For completeness here we summarize the four mean-field equations in Fermi units. The mean-field equation for the gap $\Delta$ [defined in Eq. (40)] is

$$\frac{\Delta}{g} = \Delta \sum_k G_{0,\downarrow}(-k + Q)G_\uparrow(k) = \Delta \sum_{\mathbf{k}} \frac{X_{\mathbf{k}}}{2E_{\mathbf{kQ}}}.$$ (27)

Regularizing this equation using $m/(4\pi a) = -1/g + \sum_{\mathbf{k}}(1/2\epsilon_{\mathbf{k}})$, and converting it into dimensionless form, then gives

$$0 = \Delta \left\{ (k_F a)^{-1} + \frac{1}{\pi} \int_0^\infty dx\ x^2 \int_0^\pi d\theta\ \sin(\theta) \left[ \frac{\widetilde{X}_{\mathbf{k}}}{\widetilde{E}_{\mathbf{kQ}}} - \frac{1}{\widetilde{\epsilon}_{\mathbf{k}}} \right] \right\}. \tag{28}$$

Here we defined $x = k/k_F$, $\widetilde{E}_{\mathbf{kQ}} = E_{\mathbf{kQ}}/E_F$, $\widetilde{h}_{\mathbf{kQ}} = h_{\mathbf{kQ}}/E_F$, $\widetilde{\epsilon}_{\mathbf{k}} = \epsilon_{\mathbf{k}}/E_F$, where $E_F$ is the Fermi energy and $k_F$ is the Fermi wave vector. All expressions with a tilde (such as $\widetilde{X}_{\mathbf{k}}$) are to be interpreted as having their dependent variables normalized to Fermi units.

The mean-field equation for the FF pairing vector $Q$ [defined in Eq. (78)] is

$$0 = \sum_\sigma \sum_k \left( \frac{k + Q/2}{m} \right)^z G_\sigma(k + Q/2) = \sum_{\mathbf{k}} \frac{Q/2}{m} \left[ 1 - \frac{\xi_{\mathbf{kQ}}}{E_{\mathbf{kQ}}} X_{\mathbf{k}} + \frac{k_z}{Q/2} W_{\mathbf{k}} \right]. \tag{29}$$

In terms of dimensionless variables, this equation becomes

$$0 = \widetilde{Q} \int_0^\infty dx\ x^2 \int_0^\pi d\theta\ \sin(\theta) \left[ 1 - \frac{\widetilde{\xi}_{\mathbf{kQ}}}{\widetilde{E}_{\mathbf{kQ}}} \widetilde{X}_{\mathbf{k}} + \frac{\widetilde{k}_z}{\widetilde{Q}/2} \widetilde{W}_{\mathbf{k}} \right]. \tag{30}$$

The number equations are

$$n_\sigma = \sum_k G_\sigma(k + Q/2) = \frac{1}{2} \sum_{\mathbf{k}} \left\{ [1 + f(x_{1,\sigma}) - f(x_{2,\sigma})] - \frac{\xi_{\mathbf{kQ}}}{E_{\mathbf{kQ}}} [1 - f(x_{1,\sigma}) - f(x_{2,\sigma})] \right\}. \tag{31}$$

In terms of dimensionless variables these become

$$\frac{n_\sigma}{n} = \frac{3}{8} \int_0^\infty dx\ x^2 \int_0^\pi d\theta\ \sin(\theta) \left\{ [1 + f(x_{1,\sigma}) - f(x_{2,\sigma})] - \frac{\widetilde{\xi}_{\mathbf{kQ}}}{\widetilde{E}_{\mathbf{kQ}}} [1 - f(x_{1,\sigma}) - f(x_{2,\sigma})] \right\}. \tag{32}$$

Here we used $k_F^3 = 3\pi^2 n$ as the relation between particle number and the Fermi wave vector. The two number equations can be rewritten as $n = \sum_\sigma n_\sigma$ and $p = (n_\uparrow - n_\downarrow)/(n_\uparrow + n_\downarrow)$. By taking the sum and difference of Eq. (32), we can then express these as two number equations:

$$\frac{4}{3} = \int_0^\infty dx\ x^2 \int_0^\pi d\theta\ \sin(\theta) \left[ 1 - \frac{\widetilde{\xi}_{\mathbf{kQ}}}{\widetilde{E}_{\mathbf{kQ}}} \widetilde{X}_{\mathbf{k}} \right], \tag{33}$$

$$\frac{4}{3} p = \int_0^\infty dx\ x^2 \int_0^\pi d\theta\ \sin(\theta) \widetilde{W}_{\mathbf{k}}. \tag{34}$$

There are now four equations: Eq. (28), Eq. (30), Eq. (33), and Eq. (34) in four unknowns: $\mu, \Delta, h, Q$. Thus for fixed scattering length $(1/k_F a)$ and fixed polarization $(p)$ we have a well-defined system of equations. To obtain the correct ground state, the thermodynamic potential in Eq. (91) must also be minimized. Hence one can solve the mean-field equations, and then restrict the region of parameter space to those that minimize $\Omega$.

## II. FULL VERTEX AND WARD-TAKAHASHI IDENTITY

In order to determine the full vertex and response functions in the superfluid phase, we apply the Ward-Takahashi identity (WTI). The Ward-Takahashi identity for the full vertex $\Gamma_\sigma^\mu(k_+, k_-)$ is [3]

$$\begin{aligned} q_\mu \Gamma_\sigma^\mu(k_+, k_-) &= G_\sigma^{-1}(k_+) - G_\sigma^{-1}(k_-), \\ &= q_\mu \gamma_\sigma^\mu(k_+, k_-) + \Sigma_\sigma(k_-) - \Sigma_\sigma(k_+). \end{aligned} \tag{35}$$

Here $k_\pm^\mu = k^\mu \pm q^\mu/2$, where $q^\mu = (i\Omega_m, \mathbf{q})$ with $i\Omega_m$ being a bosonic Matsubara frequency. This is an exact relation in quantum field theory which relates the single particle Green's function to the full vertex. It is a gauge invariant statement and here it reflects the underlying global $U(1)$ gauge symmetry. The bare WTI identity, $q_\mu \gamma_\sigma^\mu(k_+, k_-) = G_{0,\sigma}^{-1}(k_+) - G_{0,\sigma}^{-1}(k_-)$, is satisfied by the bare vertex $\gamma_\sigma^\mu(k_+, k_-) = \left(1, \frac{\mathbf{k}}{m}\right)$. In order to satisfy the WTI one must insert all possible vertices into the self energy Feynman diagram [3, 4]. In Fig. (1) the self energy Feynman diagram for Eq. (5) is shown.

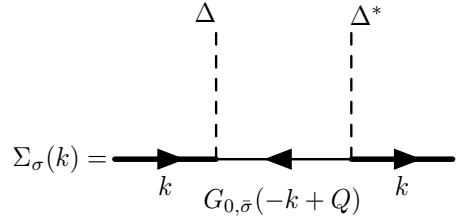

FIG. 1. Feynman diagram for the self energy $\Sigma_\sigma(k) = -|\Delta|^2 G_{0,\bar\sigma}(-k+Q)$. Note that, the external lines for $\Delta$ and $\Delta^*$ are merely for illustration. They carry no momentum, and so the self energy is indeed a two point function.

For the self energy Feynman diagram of Fig. (1) there are three possible positions where vertices can be inserted: the bare Green's function and the two gaps $\Delta$ and $\Delta^*$. Inserting current on the bare Green's function gives rise to a bare vertex. In the superfluid phase there are collective mode vertices $\Pi^\mu(q)$ and $\bar\Pi^\mu(q)$ due to the spontaneously broken global $U(1)$ gauge symmetry. Thus inserting on the gaps $\Delta$ and $\Delta^*$ leads to collective mode vertices. The full vertex is thus

$$
\begin{aligned}
\Gamma^\mu_\sigma(k_+, k_-) = {} & \gamma^\mu_\sigma(k_+, k_-) \\
& - \left[ \Delta^* \Pi^\mu(q) G_{0,\bar\sigma}(-k_- + Q) + \Delta \bar\Pi^\mu(q) G_{0,\bar\sigma}(-k_+ + Q) \right] \\
& - |\Delta|^2 G_{0,\bar\sigma}(-k_- + Q) \gamma^\mu_{\bar\sigma}(-k_- + Q, -k_+ + Q) G_{0,\bar\sigma}(-k_+ + Q).
\end{aligned}
\tag{36}
$$

The Feynman diagrams for the full vertex are expressed in Fig. (2).

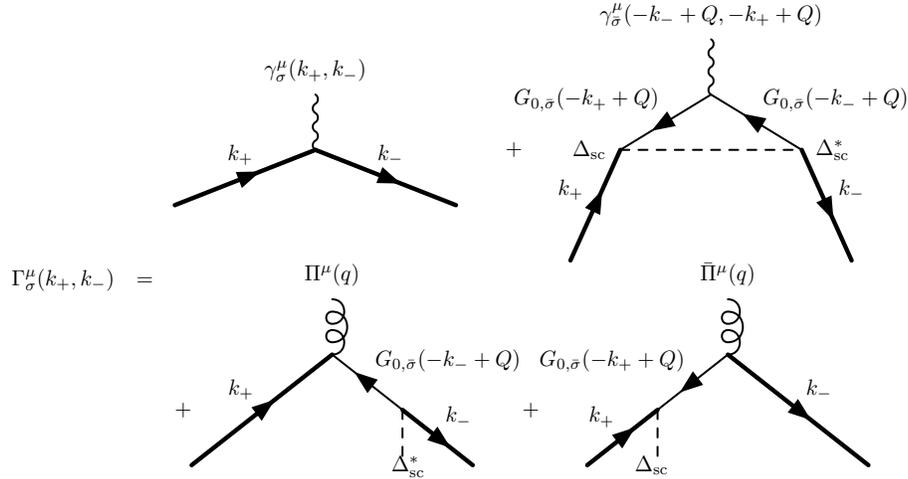

FIG. 2. Feynman diagrams for the full vertex $\Gamma^\mu_\sigma(k_+, k_-)$.

In the next section we shall verify that, for $q^\mu \neq 0$, the collective mode vertices satisfy $q_\mu \Pi^\mu(q) = 2\Delta$ and $q_\mu \bar\Pi^\mu(q) = -2\Delta^*$. Given these relations, we now verify that the WTI is indeed satisfied by the full vertex appearing in Eq. (36). Taking the contraction of the full vertex gives

$$
\begin{aligned}
q_\mu \Gamma^\mu_\sigma(k_+, k_-) = {} & G^{-1}_{0,\sigma}(k_+) - G^{-1}_{0,\sigma}(k_-) \\
& - 2|\Delta|^2 \left[ G_{0,\bar\sigma}(-k_- + Q) - G_{0,\bar\sigma}(-k_+ + Q) \right] \\
& - |\Delta|^2 G_{0,\bar\sigma}(-k_- + Q) \left[ G^{-1}_{0,\bar\sigma}(-k_- + Q) - G^{-1}_{0,\bar\sigma}(-k_+ + Q) \right] G_{0,\bar\sigma}(-k_+ + Q).
\end{aligned}
\tag{37}
$$

Collecting terms we then have

$$
\begin{aligned}
q_\mu \Gamma^\mu_\sigma(k_+, k_-) = {} & G^{-1}_{0,\sigma}(k_+) - G^{-1}_{0,\sigma}(k_-) + 2\left(\Sigma_\sigma(k_-) - \Sigma_\sigma(k_+)\right), \\
& - |\Delta|^2 \left[ G_{0,\bar\sigma}(-k_+ + Q) - G_{0,\bar\sigma}(-k_- + Q) \right], \\
= {} & G^{-1}_{0,\sigma}(k_+) - G^{-1}_{0,\sigma}(k_-) + 2\left(\Sigma_\sigma(k_-) - \Sigma_\sigma(k_+)\right) - \left(\Sigma_\sigma(k_-) - \Sigma_\sigma(k_+)\right), \\
= {} & G^{-1}_\sigma(k_+) - G^{-1}_\sigma(k_-).
\end{aligned}
\tag{38}
$$

Thus the Ward-Takahashi identity is indeed satisfied by the full vertex in Eq. (36). The full response functions are then given by

$$P^{\mu\nu}(q) = \sum_\sigma \sum_k G_\sigma(k_+) \Gamma_\sigma^\mu(k_+, k_-) G_\sigma(k_-) \gamma_\sigma^\nu(k_-, k_+). \tag{39}$$

## III. GAP EQUATION AND COLLECTIVE MODE VERTICES

In this section the explicit form of the collective mode vertices is determined. To do this, it is crucial to ensure that they are consistent with the mean-field gap equation. The gap equation is given by [5]

$$\frac{\Delta}{g} = \sum_k \mathcal{G}_{12}(k) = \sum_k \Delta G_{0,\downarrow}(-k+Q) G_\uparrow(k). \tag{40}$$

Here $g > 0$ is an $s$-wave interaction constant. This equation is equivalent to the statement that the mean-field thermodynamic potential is stationary with respect to $\Delta$. In Fig. (3) the gap equation is expressed as a Feynman diagram.

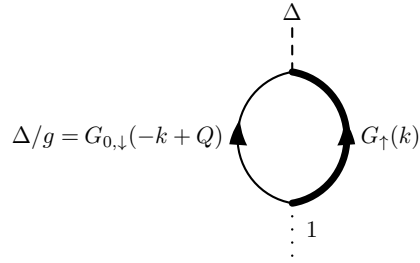

FIG. 3. Feynman diagram for the gap equation $\Delta/g = \sum_k \Delta G_{0,\downarrow}(-k+Q) G_\uparrow(k)$.

To derive the collective mode vertices, we perform all possible vertex insertions on the gap equation. In Fig. (3) there are three possible vertex insertions: a collective mode vertex can be inserted on the gap $\Delta$, a bare vertex can be inserted in the bare Green's function, and a full vertex can be inserted in the full Green's function. After performing all these possible vertex insertions on the gap equation, we obtain the Feynman diagrams in Fig. (4).

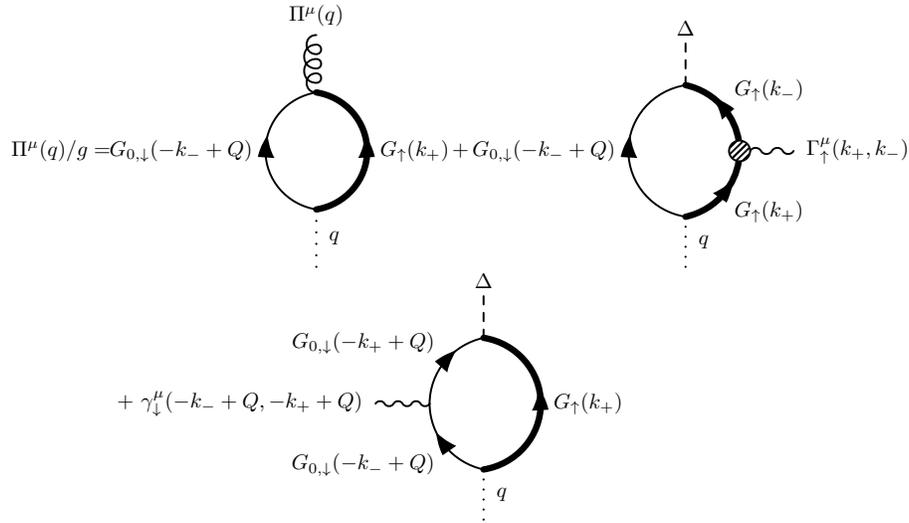

FIG. 4. Self-consistent equation for collective mode vertices after performing all possible vertex insertions on the gap equation.

Mathematically these Feynman diagrams are expressed as

$$
\begin{aligned}
\Pi^\mu/g = \Pi^\mu &\sum_k G_{0,\downarrow}(-k+Q)G_\uparrow(k+q) \\
&+ \Delta \sum_k G_{0,\downarrow}(-k+Q)G_\uparrow(k+q)\Gamma_\uparrow^\mu(k+q,k)G_\uparrow(k) \\
&+ \Delta \sum_k G_\uparrow(k+q)G_{0,\downarrow}(-k+Q)\gamma_\downarrow^\mu(-k+Q,-k-q+Q)G_{0,\downarrow}(-k-q+Q).
\end{aligned}
\tag{41}
$$

Notice that the full vertex appears in this expression. If we insert the full vertex from Eq. (36) into Eq. (41), and then shift $k \to k+Q/2$, we obtain the following expression:

$$
\begin{aligned}
\Pi^\mu &\left[1/g - \sum_k G_{0,\downarrow}(-k+Q/2)G_\uparrow(k+q+Q/2)\left(1 - |\Delta|^2 G_{0,\downarrow}(-k+Q/2)G_\uparrow(k+Q/2)\right)\right] \\
&= -\bar{\Pi}^\mu \sum_k \mathcal{G}_{12}(k+Q/2)\mathcal{G}_{12}(k+q+Q/2) \\
&+ \sum_k G_\uparrow(k+q+Q/2)\gamma_\uparrow^\mu(k+q+Q/2,k+Q/2)\mathcal{G}_{12}(k+Q/2) \\
&+ \sum_k G_{0,\downarrow}(-k+Q/2)\gamma_\downarrow^\mu(-k+Q/2,-k-q+Q/2)\mathcal{G}_{12}(k+q+Q/2)\left(1 - |\Delta|^2 G_{0,\downarrow}(-k+Q/2)G_\uparrow(k+Q/2)\right).
\end{aligned}
\tag{42}
$$

Using the property derived in Eq. (7), along with Eq. (4) and Eq. (5), we obtain

$$
\left(1 - |\Delta|^2 G_{0,\downarrow}(-k+Q/2)G_\uparrow(k+Q/2)\right) = G_\downarrow(-k+Q/2)/G_{0,\downarrow}(-k+Q/2).
\tag{43}
$$

Therefore Eq. (42) can be simplified to

$$
\begin{aligned}
\Pi^\mu &\left[1/g - \sum_k G_\downarrow(-k+Q/2)G_\uparrow(k+q+Q/2)\right] + \bar{\Pi}^\mu \sum_k \mathcal{G}_{12}(k+Q/2)\mathcal{G}_{12}(k+q+Q/2) \\
&= \sum_\sigma \sum_k G_\sigma(k+q+Q/2)\gamma_\sigma^\mu(k+q+Q/2,k+Q/2)\Delta G_{0,\bar\sigma}(-k+Q/2)G_\sigma(k+Q/2).
\end{aligned}
\tag{44}
$$

Performing the same analysis on the conjugate gap equation, $\Delta^*/g = \sum_k \Delta^* G_{0,\downarrow}(-k+Q)G_\uparrow(k)$, leads to

$$
\begin{aligned}
\bar{\Pi}^\mu &\left[1/g - \sum_k G_\downarrow(-k-q+Q/2)G_\uparrow(k+Q/2)\right] + \Pi^\mu \sum_k \mathcal{G}_{12}^*(-k-Q/2)\mathcal{G}_{12}^*(-k-q-Q/2) \\
&= \sum_\sigma \sum_k \Delta^* G_{0,\bar\sigma}(-k-q+Q/2)G_\sigma(k+q+Q/2)\gamma_\sigma^\mu(k+q+Q/2,k+Q/2)G_\sigma(k+Q/2).
\end{aligned}
\tag{45}
$$

The collective mode vertices can be written as the following matrix equation (for $q \neq 0$):

$$
\begin{pmatrix} \Pi^\mu \\ \bar{\Pi}^\mu \end{pmatrix} = \begin{pmatrix} M_{+-} & M_{++} \\ M_{--} & M_{-+} \end{pmatrix}^{-1} \begin{pmatrix} P_+^\mu \\ P_-^\mu \end{pmatrix}.
\tag{46}
$$

Here the response functions entering into the above equation are

$$
M_{+-}(q) = 1/g - \sum_k G_\uparrow(k+q+Q/2)G_\downarrow(-k+Q/2),
\tag{47}
$$

$$
M_{-+}(q) = 1/g - \sum_k G_\uparrow(k+Q/2)G_\downarrow(-k-q+Q/2),
\tag{48}
$$

$$
M_{++}(q) = \sum_k \mathcal{G}_{12}(k+Q/2)\mathcal{G}_{12}(k+q+Q/2),
\tag{49}
$$

$$
M_{--}(q) = \sum_k \mathcal{G}_{12}^*(-k-Q/2)\mathcal{G}_{12}^*(-k-q-Q/2),
\tag{50}
$$

$$
P_+^\mu(q) = \sum_\sigma \sum_k G_\sigma(k+q+Q/2)\gamma_\sigma^\mu(k+q+Q/2,k+Q/2)\Delta G_{0,\bar\sigma}(-k+Q/2)G_\sigma(k+Q/2),
\tag{51}
$$

$$
P_-^\mu(q) = \sum_\sigma \sum_k \Delta^* G_{0,\bar\sigma}(-k-q+Q/2)G_\sigma(k+q+Q/2)\gamma_\sigma^\mu(k+q+Q/2,k+Q/2)G_\sigma(k+Q/2).
\tag{52}
$$

Note that, Eq. (46) cannot be inverted at $q^\mu = 0$. This is because the poles of the collective mode vertices have a singularity associated with the Nambu-Goldstone mode that restores the global $U(1)$ gauge symmetry. It will prove convenient later to study the collective modes at $q^\mu = 0$. At $q^\mu = 0$, the collective modes obey

$$\begin{pmatrix} M_{+-}(0) & M_{++}(0) \\ M_{--}(0) & M_{-+}(0) \end{pmatrix} \begin{pmatrix} \Pi^\mu(0) \\ \bar\Pi^\mu(0) \end{pmatrix} = \begin{pmatrix} P_+^\mu(0) \\ P_-^\mu(0) \end{pmatrix}. \tag{53}$$

Inserting $q^\mu = 0$ into Eqs. (47-52), we find that $P_+^\mu(0)/\Delta = P_-^\mu(0)/\Delta^* = P_0^\mu$, and $M_{++}(0)/\Delta^2 = M_{+-}(0)/|\Delta|^2 = M_{-+}(0)/|\Delta|^2 = M_{--}(0)/(\Delta^*)^2 = M_0$, where $P_0^\mu$ and $M_0$ are defined by

$$P_0^\mu = \sum_\sigma \sum_k G_\sigma(k+Q/2)\gamma_\sigma^\mu(k+Q/2,k+Q/2)G_{0,\bar\sigma}(-k+Q/2)G_\sigma(k+Q/2), \tag{54}$$

$$M_0 = \sum_k G_{0,\downarrow}^2(-k+Q/2)G_\uparrow^2(k+Q/2). \tag{55}$$

Explicit calculation shows that $P_0^x = P_0^y = 0$. For the $z$-component, Eq. (53) gives the following equation

$$M_0\left(\Delta^*\Pi^z(0) + \Delta\bar\Pi^z(0)\right)\begin{pmatrix} \Delta \\ \Delta^* \end{pmatrix} = P_0^z\begin{pmatrix} \Delta \\ \Delta^* \end{pmatrix}. \tag{56}$$

For $\Delta \neq 0$, this equation then implies that

$$\Delta^*\Pi^z(0) + \Delta\bar\Pi^z(0) = \frac{P_0^z}{M_0}. \tag{57}$$

Returning to the general $q^\mu \neq 0$ case, the final task is to check the property of the collective mode vertices: $q_\mu\Pi^\mu(q) = 2\Delta$ and $q_\mu\bar\Pi^\mu(q) = -2\Delta^*$, which was used in verifying the WTI for the full vertex. To do this we contract each side of Eq. (46) with $q_\mu$. In order to calculate the right-hand side, we calculate the contraction $q_\mu P_\pm^\mu(q)$: explicit calculation shows that

$$q_\mu P_+^\mu(q) = 2\left(\Delta M_{+-} - \Delta^* M_{++}\right). \tag{58}$$

Similarly, since $\Delta^* P_+^\mu(q) = \Delta P_-^\mu(-q)$, we also find $q_\mu P_-^\mu(q) = -\left(q_\mu P_+^\mu(q)\right)^*$. The contractions of the collective mode vertices are then

$$\begin{pmatrix} q_\mu\Pi^\mu \\ q_\mu\bar\Pi^\mu \end{pmatrix} = \begin{pmatrix} M_{+-} & M_{++} \\ M_{--} & M_{-+} \end{pmatrix}^{-1} \begin{pmatrix} 2\left(\Delta M_{+-} - \Delta^* M_{++}\right) \\ -2\left(\Delta^* M_{-+} - \Delta M_{--}\right) \end{pmatrix} = \begin{pmatrix} 2\Delta \\ -2\Delta^* \end{pmatrix}. \tag{59}$$

This confirms that, for all $q^\mu \neq 0$, we have the desired relations

$$q_\mu\Pi^\mu(q) = 2\Delta, \quad q_\mu\bar\Pi^\mu(q) = -2\Delta^*. \tag{60}$$

## IV. SUPERFLUID DENSITY DERIVATION VIA KUBO FORMULA

### A. Kubo formulae analysis

This section derives the explicit formula for the superfluid density tensor derived in Eq. (8) of the main text. The superfluid density tensor is defined by

$$\left(n_s^{ij}/m\right) = (n/m)\,\delta^{ij} + P^{ij}(\omega = 0, \mathbf{q} \to 0). \tag{61}$$

The particle number is

$$n/m = (1/m)\sum_\sigma\sum_k G_\sigma(k). \tag{62}$$

It is convenient to perform an integration by parts on the above expression as follows:

$$
\begin{aligned}
(n/m)\,\delta^{ij} &= -\sum_\sigma \sum_k \gamma_\sigma^j(k,k)\frac{d}{dk^i}G_\sigma(k), \\
&= \sum_\sigma \sum_k \gamma_\sigma^j(k,k)G_\sigma^2(k)\frac{d}{dk^i}G_\sigma^{-1}(k), \\
&= \sum_\sigma \sum_k \gamma_\sigma^j(k,k)G_\sigma^2(k)\frac{d}{dk^i}\left(G_{0,\sigma}^{-1}(k)-\Sigma_\sigma(k)\right), \\
&= -\sum_\sigma \sum_k \gamma_\sigma^j(k,k)G_\sigma^2(k)\left(\gamma_\sigma^i(k,k)+|\Delta|^2 G_{0,\bar\sigma}^2(-k+Q)\gamma_{\bar\sigma}^i(-k+Q,-k+Q)\right).
\end{aligned}
\tag{63}
$$

Combining this expression with that for the response function in Eq. (39), and using Eq. (7), we then obtain

$$
\begin{aligned}
\left(\frac{n_s^{ij}}{m}\right) &= 4\sum_k \mathcal{G}_{12}(k+Q/2)\left(\frac{k^i-Q/2\delta^{iz}}{m}\right)\mathcal{G}_{12}^*(-k-Q/2)\left(\frac{k^j+Q/2\delta^{jz}}{m}\right) \\
&\quad -\lim_{q^k\to 0,\,q^i=q^j=0,\,\omega=0}\left[\Pi^i(q)P_-^j(-q)+\bar\Pi^i(q)P_+^j(-q)\right].
\end{aligned}
\tag{64}
$$

The first term represents a "bubble" contribution whereas the second terms represent the collective mode contribution. Here the order of limits is crucial. For instance, to compute $n_s^{ij}$, first set $\omega=0$ and $q^i=q^j=0$, then finally take $q^k\to 0$, where $k\neq i,k\neq j$.

Using the Gorkov function derived in Eq. (22), and then performing the Matsubara frequency summation in the bubble term in Eq. (64), the superfluid density becomes

$$
\begin{aligned}
\left(\frac{n_s^{ij}}{m}\right) &= \sum_{\mathbf{k}} \frac{|\Delta|^2}{E_{\mathbf{kQ}}^2}\left(\frac{X_{\mathbf{k}}}{E_{\mathbf{k}}}-\beta Y_{\mathbf{k}}\right)\left(\frac{k^i-Q/2\delta^{iz}}{m}\right)\left(\frac{k^j+Q/2\delta^{jz}}{m}\right) \\
&\quad -\lim_{q^k\to 0,\,q^i=q^j=0,\,\omega=0}\left[\Pi^i(q)P_-^j(-q)+\bar\Pi^i(q)P_+^j(-q)\right].
\end{aligned}
\tag{65}
$$

As noted previously, $P_0^x=P_0^y=0$. By taking the limit in Eq. (65) in the appropriate order, and using Eq. (57), the superfluid density then becomes

$$
\left(\frac{n_s^{ij}}{m}\right) = \sum_{\mathbf{k}} \frac{|\Delta|^2}{E_{\mathbf{kQ}}^2}\left(\frac{X_{\mathbf{k}}}{E_{\mathbf{k}}}-\beta Y_{\mathbf{k}}\right)\left(\frac{k^i-Q/2\delta^{iz}}{m}\right)\left(\frac{k^j+Q/2\delta^{jz}}{m}\right)-\delta^{iz}\delta^{jz}\frac{(P_0^z)^2}{M_0}.
\tag{66}
$$

This produces Eq. (8) of the main text. Note that, when $\Delta=0$, $Q=0$ is a solution to the mean-field equations and thus $P_0^z=0$, so that the above expression does indeed vanish in the normal state. Similarly, for a non-FF superfluid, where $Q=0$, $P_0^z=0$ so that the above expression also reduces to the known result [6] in this limit. For a complete expression for the superfluid density, it only remains to compute $P_0^z$ and $M_0$. The definition of these quantities appears in Eq. (54). Evaluating those expressions gives

$$
P_0^z = \sum_{\mathbf{k}} \frac{1}{2E_{\mathbf{kQ}}}\left[\beta Z_{\mathbf{k}}\frac{k^z}{m}+\frac{\xi_{\mathbf{kQ}}}{E_{\mathbf{kQ}}}\left(\beta Y_{\mathbf{k}}-\frac{X_{\mathbf{k}}}{E_{\mathbf{kQ}}}\right)\frac{Q/2}{m}\right],
\tag{67}
$$

$$
M_0 = \sum_{\mathbf{k}} \frac{1}{4E_{\mathbf{kQ}}^2}\left(\frac{X_{\mathbf{k}}}{E_{\mathbf{kQ}}}-\beta Y_{\mathbf{k}}\right).
\tag{68}
$$

The bubble term in Eq. (66) is in agreement with Eq. (16) of [2], as we will show in Sec. (IV B). However, our complete expression has the inclusion of collective modes.

### B.  Comparison of bubble term with the literature

In this section we compare our expression for the superfluid density with that appearing in Ref. [2]. The superfluid density is defined in Eq. (61). Performing the Matsubara frequency summation in Eq. (62) gives the particle number as follows

$$
\left(\frac{n}{m}\right)\delta^{ij} = \sum_{\mathbf{k}}\left(1-\frac{\xi_{\mathbf{kQ}}}{E_{\mathbf{kQ}}}X_{\mathbf{k}}\right)\frac{\delta^{ij}}{m}.
\tag{69}
$$

Using Eq. (39), and performing the Matsubara frequency summation, for $P^{ij}(0)$ we find

$$P^{ij}(0) = -\sum_{\mathbf{k}} \left\{ \left( \frac{k^i}{m} \right)^2 \delta^{ij} \beta Y_{\mathbf{k}} + 2 \frac{k_z}{m} \frac{Q/2}{m} \delta^{iz} \delta^{jz} \beta \frac{\xi_{\mathbf{kQ}}}{E_{\mathbf{kQ}}} Z_{\mathbf{k}} + \left( \frac{Q/2}{m} \right)^2 \delta^{iz} \delta^{jz} \left[ \left( \frac{\xi_{\mathbf{kQ}}}{E_{\mathbf{kQ}}} \right)^2 \beta Y_{\mathbf{k}} + \frac{\Delta^2}{E_{\mathbf{kQ}}^2} \frac{X_{\mathbf{k}}}{E_{\mathbf{kQ}}} \right] \right\}$$
$$- \delta^{iz} \delta^{jz} \frac{(P_0^z)^2}{M_0}. \tag{70}$$

Combining Eq. (69) and Eq. (70) then gives the superfluid density

$$\left( \frac{n_s^{ij}}{m} \right) = \sum_{\mathbf{k}} \left( 1 - \frac{\xi_{\mathbf{kQ}}}{E_{\mathbf{kQ}}} X_{\mathbf{k}} \right) \frac{\delta^{ij}}{m} - \sum_{\mathbf{k}} \left\{ \left( \frac{k^i}{m} \right)^2 \delta^{ij} \beta Y_{\mathbf{k}} + 2 \frac{k_z}{m} \frac{Q/2}{m} \delta^{iz} \delta^{jz} \beta \frac{\xi_{\mathbf{kQ}}}{E_{\mathbf{kQ}}} Z_{\mathbf{k}} \right.$$
$$\left. + \left( \frac{Q/2}{m} \right)^2 \delta^{iz} \delta^{jz} \left[ \left( \frac{\xi_{\mathbf{kQ}}}{E_{\mathbf{kQ}}} \right)^2 \beta Y_{\mathbf{k}} + \frac{\Delta^2}{E_{\mathbf{kQ}}^2} \frac{X_{\mathbf{k}}}{E_{\mathbf{kQ}}} \right] \right\} - \delta^{iz} \delta^{jz} \frac{(P_0^z)^2}{M_0}. \tag{71}$$

This is in agreement with Eq. (16) of Ref. [2], up to our inclusion of collective modes, and a factor of 1/4 in the first two summations, which arises from our $Q/2$ being the $Q$ of Ref. [2]. It only remains to prove the equivalence between Eq. (66) and Eq. (71). By performing integration by parts on the number equation in Eq. (69), we have

$$\left( \frac{n}{m} \right) \delta^{ij} = \sum_{\mathbf{k}} \frac{k^j}{m} \frac{d}{dk_i} \left( \frac{\xi_{\mathbf{kQ}}}{E_{\mathbf{kQ}}} X_{\mathbf{k}} \right). \tag{72}$$

Computing the derivative then gives

$$\left( \frac{n}{m} \right) \delta^{ij} = \sum_{\mathbf{k}} \left\{ \left[ \frac{\Delta^2}{E_{\mathbf{kQ}}^2} \frac{X_{\mathbf{k}}}{E_{\mathbf{kQ}}} + \beta Y_{\mathbf{k}} \left( 1 - \frac{\Delta^2}{E_{\mathbf{kQ}}^2} \right) \right] \frac{k^i}{m} \frac{k^j}{m} + \frac{k_z}{m} \frac{Q/2}{m} \delta^{iz} \delta^{jz} \beta \frac{\xi_{\mathbf{kQ}}}{E_{\mathbf{kQ}}} Z_{\mathbf{k}} \right\}. \tag{73}$$

The superfluid density thus becomes

$$\left( \frac{n_s^{ij}}{m} \right) = \sum_{\mathbf{k}} \frac{|\Delta|^2}{E_{\mathbf{kQ}}^2} \left( \frac{X_{\mathbf{k}}}{E_{\mathbf{kQ}}} - \beta Y_{\mathbf{k}} \right) \left( \frac{k^i - Q/2\delta^{iz}}{m} \right) \left( \frac{k^j + Q/2\delta^{jz}}{m} \right)$$
$$- \sum_{\mathbf{k}} \left[ \frac{k_z}{m} \frac{Q/2}{m} \delta^{iz} \delta^{jz} \beta \frac{\xi_{\mathbf{kQ}}}{E_{\mathbf{kQ}}} Z_{\mathbf{k}} + \left( \frac{Q/2}{m} \right)^2 \delta^{iz} \delta^{jz} \beta Y_{\mathbf{k}} \right] - \delta^{iz} \delta^{jz} \frac{(P_0^z)^2}{M_0}. \tag{74}$$

Using the $W_{\mathbf{k}}$ function from Eq. (23), the superfluid density can be written as

$$\left( \frac{n_s^{ij}}{m} \right) = \sum_{\mathbf{k}} \frac{|\Delta|^2}{E_{\mathbf{kQ}}^2} \left( \frac{X_{\mathbf{k}}}{E_{\mathbf{kQ}}} - \beta Y_{\mathbf{k}} \right) \left( \frac{k^i - Q/2\delta^{iz}}{m} \right) \left( \frac{k^j + Q/2\delta^{jz}}{m} \right) + \frac{Q/2}{m} \delta^{iz} \delta^{jz} \sum_{\mathbf{k}} \frac{dW_{\mathbf{k}}}{dk_z} - \delta^{iz} \delta^{jz} \frac{(P_0^z)^2}{M_0}. \tag{75}$$

The second term gives zero contribution. Therefore the superfluid density reduces to

$$\left( \frac{n_s^{ij}}{m} \right) = \sum_{\mathbf{k}} \frac{|\Delta|^2}{E_{\mathbf{kQ}}^2} \left( \frac{X_{\mathbf{k}}}{E_{\mathbf{kQ}}} - \beta Y_{\mathbf{k}} \right) \left( \frac{k^i - Q/2\delta^{iz}}{m} \right) \left( \frac{k^j + Q/2\delta^{jz}}{m} \right) - \delta^{iz} \delta^{jz} \frac{(P_0^z)^2}{M_0}. \tag{76}$$

This proves the equivalence between Eq. (66) and Eq. (71).

## V. SUPERFLUID DENSITY DERIVATION VIA EQUILIBRIUM CURRENT

### A. Equilibrium current analysis

The equilibrium current in the $z$-direction is

$$j^z = \sum_{\sigma} \sum_{k} \left( \frac{k + Q/2}{m} \right)^z G_{\sigma}(k + Q/2). \tag{77}$$

After performing the Matsubara frequency summation, we obtain

$$j^z = \sum_{\mathbf{k}} \frac{Q/2}{m} \left[ 1 - \frac{\xi_{\mathbf{kQ}}}{E_{\mathbf{kQ}}} X_{\mathbf{k}} + \frac{k_z}{Q/2} W_{\mathbf{k}} \right], \tag{78}$$

which is in agreement with Eq. (20) of Ref. [2]. The self-consistent condition for the FF pairing vector $\mathbf{Q} = Q\hat{\mathbf{z}}$ is that $j^z = 0$. This condition is equivalent to the statement that the mean-field thermodynamic potential is stationary with respect to $Q$. In what follows we will need the derivative of Eq. (77) with respect to $Q$, with $\mu$ and $h$ fixed at their mean-field values. This will require the following lemma:

$$\frac{\partial}{\partial Q} G_\sigma^{-1}(k+Q/2) \bigg|_{\mu,h} = -\frac{1}{2} \Gamma_\sigma^z(k+Q/2, k+Q/2). \tag{79}$$

This result is proved as follows:

$$\begin{aligned}
\frac{\partial}{\partial Q} G_\sigma^{-1}(k+Q/2) \bigg|_{\mu,h} &= \frac{\partial}{\partial Q} \left( G_{0,\sigma}^{-1}(k+Q/2) - \Sigma_\sigma(k+Q/2) \right) \bigg|_{\mu,h}, \\
&= \frac{\partial}{\partial Q} \left( i\omega - \xi_{\mathbf{k+Q}/2,\sigma} + |\Delta|^2 G_{0,\bar{\sigma}}(-k+Q/2) \right) \bigg|_{\mu,h}, \\
&= -\frac{1}{2m} (k+Q/2)^z + \frac{\partial |\Delta|^2}{\partial Q} \bigg|_{\mu,h} G_{0,\bar{\sigma}}(-k+Q/2) \\
&\quad - |\Delta|^2 G_{0,\bar{\sigma}}^2(-k+Q/2) \frac{\partial}{\partial Q} G_{0,\bar{\sigma}}^{-1}(-k+Q/2) \bigg|_{\mu,h}, \\
&= -\frac{1}{2} \bigg[ \gamma_\sigma^z(k+Q/2, k+Q/2) - 2\frac{\partial |\Delta|^2}{\partial Q} \bigg|_{\mu,h} G_{0,\bar{\sigma}}(-k+Q/2) \\
&\quad - |\Delta|^2 G_{0,\bar{\sigma}}^2(-k+Q/2) \gamma_{\bar{\sigma}}^z(-k+Q/2, -k+Q/2) \bigg].
\end{aligned} \tag{80}$$

From the gap equation in Eq. (40) and the collective mode equation in Eq. (57) it follows that

$$2\frac{\partial |\Delta|^2}{\partial Q} \bigg|_{\mu,h} = \frac{P_0^z}{M_0} = \Delta^* \Pi^z(0) + \Delta \bar{\Pi}^z(0). \tag{81}$$

Inserting this relation into Eq. (80), and then using Eq. (36), gives

$$\begin{aligned}
\frac{\partial}{\partial Q} G_\sigma^{-1}(k+Q/2) \bigg|_{\mu,h} &= -\frac{1}{2} \bigg[ \gamma_\sigma^z(k+Q/2, k+Q/2) - |\Delta|^2 G_{0,\bar{\sigma}}^2(-k+Q/2) \gamma_{\bar{\sigma}}^z(-k+Q/2, -k+Q/2) \\
&\quad - \left( \Delta^* \Pi^z + \Delta \bar{\Pi}^z \right) G_{0,\bar{\sigma}}(-k+Q/2) \bigg] \\
&= -\frac{1}{2} \Gamma_\sigma^z(k+Q/2, k+Q/2).
\end{aligned} \tag{82}$$

Thus the lemma is proved.

Taking the derivative of Eq. (77), and using the lemma in Eq. (79), then gives

$$
\begin{aligned}
\frac{\partial j^z}{\partial Q}\bigg|_{\mu,h} &= \frac{\partial}{\partial Q} \sum_\sigma \sum_k \left(\frac{k+Q/2}{m}\right)^z G_\sigma(k+Q/2), \\
&= \frac{1}{2m} \sum_\sigma \sum_k G_\sigma(k+Q/2) + \sum_\sigma \sum_k \left(\frac{k+Q/2}{m}\right)^z \frac{\partial}{\partial Q} G_\sigma(k+Q/2)\bigg|_{\mu,h}, \\
&= \frac{n}{2m} - \sum_\sigma \sum_k \left(\frac{k+Q/2}{m}\right)^z G_\sigma^2(k+Q/2) \left.\frac{\partial}{\partial Q} G_\sigma^{-1}(k+Q/2)\right|_{\mu,h}, \\
&= \frac{1}{2} \left(\frac{n}{m} + \sum_\sigma \sum_k G_\sigma(k+Q/2)\Gamma_\sigma^z(k+Q/2,k+Q/2)G_\sigma(k+Q/2)\gamma_\sigma^z(k+Q/2,k+Q/2)\right), \\
&= \frac{1}{2} \left(\frac{n}{m} + P^{zz}(0)\right), \\
&= \frac{1}{2} \left(\frac{n_s^{zz}}{m}\right).
\end{aligned}
\tag{83}
$$

Thus, the $z,z$-component of the superfluid density tensor is the derivative of $j^z$ with respect to $Q$.

Taking the explicit $Q$-derivative of Eq. (78), with $\mu$ and $h$ fixed at their mean-field values, gives

$$
\frac{\partial j^z}{\partial Q}\bigg|_{\mu,h} = \sum_{\mathbf{k}} \frac{Q/2}{m} \frac{\partial}{\partial Q} \left[1 - \frac{\xi_{\mathbf{kQ}}}{E_{\mathbf{kQ}}}X_{\mathbf{k}} + \frac{k_z}{Q/2}W_{\mathbf{k}}\right]\bigg|_{\mu,h}
\tag{84}
$$

Here the saddle point equation $j^z = 0$ has been used to simplify the above expression. Performing the remaining $Q$-derivative, and accounting for the $Q$-dependence in the gap $\Delta$ through Eq. (81), we obtain

$$
\begin{aligned}
\frac{\partial j^z}{\partial Q}\bigg|_{\mu,h} &= \frac{1}{2} \sum_{\mathbf{k}} \left(1 - \frac{\xi_{\mathbf{kQ}}}{E_{\mathbf{kQ}}}X_{\mathbf{k}}\right)\frac{\delta^{ij}}{m} - \frac{1}{2} \sum_{\mathbf{k}} \left\{\left(\frac{k^i}{m}\right)^2 \delta^{ij}\beta Y_{\mathbf{k}} + 2\frac{k_z}{m}\frac{Q/2}{m}\delta^{iz}\delta^{jz}\beta\frac{\xi_{\mathbf{kQ}}}{E_{\mathbf{kQ}}}Z_{\mathbf{k}} \right. \\
&\quad \left. + \left(\frac{Q/2}{m}\right)^2 \delta^{iz}\delta^{jz}\left[\left(\frac{\xi_{\mathbf{kQ}}}{E_{\mathbf{kQ}}}\right)^2 \beta Y_{\mathbf{k}} + \frac{\Delta^2}{E_{\mathbf{kQ}}^2}\frac{X_{\mathbf{k}}}{E_{\mathbf{kQ}}}\right]\right\} - \frac{1}{2}\delta^{iz}\delta^{jz}\frac{(P_0^z)^2}{M_0}.
\end{aligned}
\tag{85}
$$

Thus, from Eq. (83), it follows that the superfluid density is

$$
\begin{aligned}
\left(\frac{n_s^{zz}}{m}\right) &= \sum_{\mathbf{k}} \left(1 - \frac{\xi_{\mathbf{kQ}}}{E_{\mathbf{kQ}}}X_{\mathbf{k}}\right)\frac{\delta^{ij}}{m} - \sum_{\mathbf{k}} \left\{\left(\frac{k^i}{m}\right)^2 \delta^{ij}\beta Y_{\mathbf{k}} + 2\frac{k_z}{m}\frac{Q/2}{m}\delta^{iz}\delta^{jz}\beta\frac{\xi_{\mathbf{kQ}}}{E_{\mathbf{kQ}}}Z_{\mathbf{k}} \right. \\
&\quad \left. + \left(\frac{Q/2}{m}\right)^2 \delta^{iz}\delta^{jz}\left[\left(\frac{\xi_{\mathbf{kQ}}}{E_{\mathbf{kQ}}}\right)^2 \beta Y_{\mathbf{k}} + \frac{\Delta^2}{E_{\mathbf{kQ}}^2}\frac{X_{\mathbf{k}}}{E_{\mathbf{kQ}}}\right]\right\} - \delta^{iz}\delta^{jz}\frac{(P_0^z)^2}{M_0}.
\end{aligned}
\tag{86}
$$

This reproduces the result in Eq. (71), which was shown in Sec. (IV B) to be equivalent to Eq. (66).

## B. Vanishing of the superfluid density in directions transverse to pairing vector

Interestingly, due to the underlying rotational invariance of the FF state [7], the perpendicular component of the superfluid density vanishes: $n_s^{xx} = n_s^{yy} = 0$. This result was proved, analytically for $T = 0$ and numerically for finite $T$, in Ref. [2]. Here we analytically prove that $n_s^{xx} = 0$ for all temperatures. Setting $i = j = x$ in Eq. (65) gives zero collective mode contribution; thus we obtain

$$
\left(\frac{n_s^{xx}}{m}\right) = \sum_{\mathbf{k}} \frac{|\Delta|^2}{E_{\mathbf{kQ}}^2}\left(\frac{X_{\mathbf{k}}}{E_{\mathbf{kQ}}} - \beta Y_{\mathbf{k}}\right)\left(\frac{k^x}{m}\right)^2.
\tag{87}
$$

The self-consistent equation for $Q$ is given in Eq. (78). After performing integration by parts on the expression in Eq. (78), we obtain

$$
j^z = -(Q/2) \sum_{\mathbf{k}} \frac{k^x}{m}\frac{d}{dk^x}\left[1 - \frac{\xi_{\mathbf{kQ}}}{E_{\mathbf{kQ}}}X_{\mathbf{k}} + \frac{k^z}{Q/2}W_{\mathbf{k}}\right].
\tag{88}
$$

Evaluating the derivatives and simplifying then gives

$$j^z = (Q/2) \left( \frac{n_s^{xx}}{m} \right) - m \sum_{\mathbf{k}} \left( \frac{k^x}{m} \right)^2 \frac{dW_{\mathbf{k}}}{dk^z}. \tag{89}$$

The second term gives zero contribution, and thus

$$\left( \frac{n_s^{xx}}{m} \right) = \frac{j^z}{Q/2} = 0. \tag{90}$$

In the last step the saddle-point condition $j^z = 0$ has been used. Thus the $x, x$-component (and similarly the $y, y$-component) of the superfluid density vanishes as a result of the self-consistent condition for the pairing vector $Q$.

## VI. THERMODYNAMIC POTENTIAL AND STABILITY CRITERIA

In Eq. (83) it was shown that the superfluid density is related to the partial derivative of the $z$-component of the equilibrium current with $\mu$ and $h$ fixed at their mean-field values. This equation can be expressed in terms of the thermodynamic potential, which then allows constraints on the stability of the FF phase to be studied. The mean-field thermodynamic potential is given by [1, 2]

$$\Omega = \frac{\Delta^2}{g} - \beta^{-1} \sum_{\mathbf{k}} \left\{ \log \left[ 2 \cosh(\beta E_{\mathbf{kQ}}) + 2 \cosh(\beta h_{\mathbf{kQ}}) \right] - \beta \xi_{\mathbf{kQ}} \right\}. \tag{91}$$

It assumed here that $\Delta = \Delta^*$. The saddle-point conditions are given by

$$\left. \frac{\partial \Omega}{\partial \Delta} \right|_{\mu, h, Q} = 0, \quad \left. \frac{\partial \Omega}{\partial Q} \right|_{\mu, h, \Delta} = 0. \tag{92}$$

The first of these conditions reproduces the gap equation

$$2 \left( \frac{\Delta}{g} - \Delta \sum_{\mathbf{k}} \frac{X_{\mathbf{k}}}{2 E_{\mathbf{kQ}}} \right) = 0, \tag{93}$$

while the second reproduces the constraint that the equilibrium current $j^z$ vanishes

$$\frac{1}{2} \sum_{\mathbf{k}} \frac{Q/2}{m} \left[ 1 - \frac{\xi_{\mathbf{kQ}}}{E_{\mathbf{kQ}}} X_{\mathbf{k}} + \frac{k_z}{Q/2} W_{\mathbf{k}} \right] = 0. \tag{94}$$

The partial derivative of $j^z$ at fixed $\mu$ and $h$ is given by

$$\left. \frac{\partial j^z}{\partial Q} \right|_{\mu, h} = \left. \frac{\partial j^z}{\partial Q} \right|_{\mu, h, \Delta} + \left. \frac{\partial j^z}{\partial \Delta} \right|_{\mu, h, Q} \left. \frac{\partial \Delta}{\partial Q} \right|_{\mu, h}. \tag{95}$$

Since the equilibrium current is related to the thermodynamic potential by $j^z = 2 \left. (\partial \Omega / \partial Q) \right|_{\mu, h, \Delta}$, Eq. (95) can be expressed as

$$\frac{1}{2} \left. \frac{\partial j^z}{\partial Q} \right|_{\mu, h} = \left. \frac{\partial^2 \Omega}{\partial Q^2} \right|_{\mu, h, \Delta} + \left[ \left. \frac{\partial}{\partial \Delta} \left( \left. \frac{\partial \Omega}{\partial Q} \right) \right|_{\mu, h, \Delta} \right] \right|_{\mu, h, Q} \left. \frac{\partial \Delta}{\partial Q} \right|_{\mu, h}. \tag{96}$$

To compute the remaining term, $\left. (\partial \Delta / \partial Q) \right|_{\mu, h}$, we first consider the gap equation:

$$\left. \frac{\partial \Omega}{\partial \Delta} \right|_{\mu, h, Q} = 0. \tag{97}$$

Differentiating this equation with respect to $Q$, at fixed $\mu$ and $h$, then gives

$$\frac{\partial}{\partial Q} \left[ \left. \frac{\partial \Omega}{\partial \Delta} \right|_{\mu, h, Q} \right] \right|_{\mu, h} = 0,$$

$$\Rightarrow \left[ \left. \frac{\partial}{\partial Q} \left( \left. \frac{\partial \Omega}{\partial \Delta} \right) \right|_{\mu, h, Q} \right] \right|_{\mu, h, \Delta} + \left. \frac{\partial^2 \Omega}{\partial \Delta^2} \right|_{\mu, h, Q} \left. \frac{\partial \Delta}{\partial Q} \right|_{\mu, h} = 0. \tag{98}$$

Rearranging then gives

$$\left. \frac{\partial \Delta}{\partial Q} \right|_{\mu, h} = - \left[ \left. \frac{\partial}{\partial Q} \left( \left. \frac{\partial \Omega}{\partial \Delta} \right) \right|_{\mu, h, Q} \right] \right|_{\mu, h, \Delta} \left/ \left. \frac{\partial^2 \Omega}{\partial \Delta^2} \right|_{\mu, h, Q} \right. . \tag{99}$$

Inserting this into Eq. (96) then gives

$$\left. \frac{1}{2} \frac{\partial j^z}{\partial Q} \right|_{\mu, h} = \left. \frac{\partial^2 \Omega}{\partial Q^2} \right|_{\mu, h, \Delta} - \left[ \left. \frac{\partial}{\partial \Delta} \left( \left. \frac{\partial \Omega}{\partial Q} \right) \right|_{\mu, h, \Delta} \right] \right|_{\mu, h, Q} \left[ \left. \frac{\partial}{\partial Q} \left( \left. \frac{\partial \Omega}{\partial \Delta} \right) \right|_{\mu, h, Q} \right] \right|_{\mu, h, \Delta} \left/ \left. \frac{\partial^2 \Omega}{\partial \Delta^2} \right|_{\mu, h, Q} \right. . \tag{100}$$

Explicit calculation shows that the second order partial derivatives are symmetric, thus this expression can be simplified to

$$\frac{1}{4} \left( \frac{n_s^{zz}}{m} \right) = \left. \frac{1}{2} \frac{\partial j^z}{\partial Q} \right|_{\mu, h} = \left. \frac{\partial^2 \Omega}{\partial Q^2} \right|_{\mu, h, \Delta} - \left\{ \left[ \left. \frac{\partial}{\partial \Delta} \left( \left. \frac{\partial \Omega}{\partial Q} \right) \right|_{\mu, h, \Delta} \right] \right|_{\mu, h, Q} \right\}^2 \left/ \left. \frac{\partial^2 \Omega}{\partial \Delta^2} \right|_{\mu, h, Q} \right. . \tag{101}$$

Thus, by taking various partial derivatives of the thermodynamic potential $\Omega$, one can compute the superfluid density in the $z, z$-direction, with collective mode contributions incorporated. The first term is the bubble contribution while the second term is the collective mode contribution. By explicit calculation, we obtain

$$\left. \frac{\partial^2 \Omega}{\partial Q^2} \right|_{\mu, h, \Delta} = \frac{1}{4} \sum_{\mathbf{k}} \left( 1 - \frac{\xi_{\mathbf{kQ}}}{E_{\mathbf{kQ}}} X_{\mathbf{k}} \right) \frac{\delta^{ij}}{m} - \frac{1}{4} \sum_{\mathbf{k}} \left\{ \left( \frac{k^i}{m} \right)^2 \delta^{ij} \beta Y_{\mathbf{k}} + 2 \frac{k_z}{m} \frac{Q/2}{m} \delta^{iz} \delta^{jz} \beta \frac{\xi_{\mathbf{kQ}}}{E_{\mathbf{kQ}}} Z_{\mathbf{k}} \right.$$
$$\left. + \left( \frac{Q/2}{m} \right)^2 \delta^{iz} \delta^{jz} \left[ \left( \frac{\xi_{\mathbf{kQ}}}{E_{\mathbf{kQ}}} \right)^2 \beta Y_{\mathbf{k}} + \frac{\Delta^2}{E_{\mathbf{kQ}}^2} \frac{X_{\mathbf{k}}}{E_{\mathbf{kQ}}} \right] \right\}, \tag{102}$$

$$\left[ \left. \frac{\partial}{\partial \Delta} \left( \left. \frac{\partial \Omega}{\partial Q} \right) \right|_{\mu, h, \Delta} \right] \right|_{\mu, h, Q} = -\Delta P_0^z, \tag{103}$$

$$\left. \frac{\partial^2 \Omega}{\partial \Delta^2} \right|_{\mu, h, Q} = 4 \Delta^2 M_0, \tag{104}$$

where $P_0^z$ and $M_0$ are defined in Eq. (67). Inserting these definitions into Eq. (101) and simplifying then reproduces

$$\left( \frac{n_s^{zz}}{m} \right) = \sum_{\mathbf{k}} \left( 1 - \frac{\xi_{\mathbf{kQ}}}{E_{\mathbf{kQ}}} X_{\mathbf{k}} \right) \frac{\delta^{ij}}{m} - \sum_{\mathbf{k}} \left\{ \left( \frac{k^i}{m} \right)^2 \delta^{ij} \beta Y_{\mathbf{k}} + 2 \frac{k_z}{m} \frac{Q/2}{m} \delta^{iz} \delta^{jz} \beta \frac{\xi_{\mathbf{kQ}}}{E_{\mathbf{kQ}}} Z_{\mathbf{k}} \right.$$
$$\left. + \left( \frac{Q/2}{m} \right)^2 \delta^{iz} \delta^{jz} \left[ \left( \frac{\xi_{\mathbf{kQ}}}{E_{\mathbf{kQ}}} \right)^2 \beta Y_{\mathbf{k}} + \frac{\Delta^2}{E_{\mathbf{kQ}}^2} \frac{X_{\mathbf{k}}}{E_{\mathbf{kQ}}} \right] \right\} - \delta^{iz} \delta^{jz} \frac{(P_0^z)^2}{M_0}, \tag{105}$$

which is the expression for $n_s^{zz}$ obtained in Eq. (71) and is equivalent to Eq. (66).

The stability criteria for the FF superfluid are [5]

$$\left. \frac{\partial^2 \Omega}{\partial Q^2} \right|_{\mu, h, \Delta} - \left\{ \left[ \left. \frac{\partial}{\partial \Delta} \left( \left. \frac{\partial \Omega}{\partial Q} \right) \right|_{\mu, h, \Delta} \right] \right|_{\mu, h, Q} \right\}^2 \left/ \left. \frac{\partial^2 \Omega}{\partial \Delta^2} \right|_{\mu, h, Q} \right. > 0, \tag{106}$$

$$\left. \frac{\partial^2 \Omega}{\partial \Delta^2} \right|_{\mu, h, Q} > 0. \tag{107}$$

The first expression is related to $n_s^{zz}$ by Eq. (101). Thus the FF phase is stable if the following two (sufficient) conditions are satisfied:

$$n_s^{zz} > 0, \tag{108}$$

$$\left. \frac{\partial^2 \Omega}{\partial \Delta^2} \right|_{\mu, h, Q} > 0. \tag{109}$$



\end{document}